\documentclass[10pt,conference,twocolumn]{IEEEtran}

\title{On the Block Error Probability \\
       of LP Decoding of LDPC Codes}
 \author{
\authorblockN{Ralf Koetter}
\authorblockA{CSL and Dept.~of ECE \\
              University of Illinois at Urbana-Champaign \\
              Urbana, IL 61801, USA \\
              \texttt{koetter@uiuc.edu}}
\and
\authorblockN{Pascal O.~Vontobel}
\authorblockA{Dept.~of EECS \\
              Massachusetts Institute of Technology \\
              Cambridge, MA 02139, USA \\
              \texttt{pascal.vontobel@ieee.org}}
}

\usepackage{amssymb, amsmath}
\usepackage{epsfig,bbm}

\def\F{{\mathbb{F}}}

\def\del{{\partial}}
\newcommand{\R}{\mathbb{R}}

\newcommand{\matr}[1]{\mathbf{#1}}
\newcommand{\vect}[1]{\mathbf{#1}}
\newcommand{\set}[1]{\mathcal{#1}}
\newcommand{\code}[1]{\mathcal{#1}}
\newcommand{\graph}[1]{\mathsf{#1}}

\newcommand{\vb}{\vect{b}}

\newcommand{\vx}{\vect{x}}

\newcommand{\vy}{\vect{y}}

\newcommand{\vzero}{\vect{0}}
\newcommand{\zero}{\vzero}

\newcommand{\vlambda}{\boldsymbol{\lambda}}
\newcommand{\vtau}{\boldsymbol{\tau}}

\newcommand{\valpha}{\boldsymbol{\alpha}}
\newcommand{\vbeta}{\boldsymbol{\beta}}

\newcommand{\mut}[1]{\mu^{(#1)}}

\newcommand{\nut}[1]{\nu^{(#1)}}

\newcommand{\vrho}{\boldsymbol{\rho}}

\newcommand{\fph}[2]{\mathcal{#1}(\matr{#2})}

\newcommand{\defeq}{\triangleq}

\newcommand{\tr}{\mathsf{T}}

\newcommand{\neighborhood}[1]{\del(#1)}

\newcommand{\Vv}{\set{V}_{\mathrm{v}}}
\newcommand{\Vc}{\set{V}_{\mathrm{c}}}

\newcommand{\convhull}{\operatorname{conv}}

\newcommand{\GF}[1]{\mathbb{F}_{#1}}

\newtheorem{Definition}{Definition}

\newtheorem{Lemma}[Definition]{Lemma}
\newtheorem{Theorem}[Definition]{Theorem}

\newenvironment{Proof}%
  {\noindent \emph{Proof:}}{\hfill$\square$}

\newcommand{\progbox}[2]
{
  \vspace{3mm}
  \noindent\fbox{%\rule[-0.15cm]{0.15cm}{2.20cm}\ \
  \begin{minipage}[b]{3.4in}{{\underline{\bf{#1}}}}

  #2
  \end{minipage}}
  \vspace{0mm}
}

\begin{document}

\maketitle

\begin{abstract}
  In his thesis, Wiberg showed the existence of thresholds for families of
  regular low-density parity-check codes under min-sum algorithm decoding. He
  also derived analytic bounds on these thresholds. In this paper, we
  formulate similar results for linear programming decoding of regular
  low-density parity-check codes.
\end{abstract}

\section{Introduction}

The goal of this paper is to shed some light on the connection between min-sum
algorithm (MSA) decoding and the formulation of decoding as a linear
program. In particular, we address the problem of bounding the performance of
linear programming (LP) decoding with respect to word error rate. The bounds
reflect similar analytic bounds for MSA decoding of low-density parity-check
(LDPC) codes due to Wiberg~\cite{Wiberg:96} and establish the existence of an
SNR threshold for LP decoding. While highly efficient and structured
computer-based evaluation techniques, such as density evolution (see
e.g.~\cite{Luby:Mitzenmacher:Shokrollahi:Spielman:Stemann:97:1,
Luby:Mitzenmacher:Shokrollahi:98:1, Richardson:Urbanke:01:2,
Lentmaier:Truhachev:Costello:Zigangirov:04:1, Jin:Richardson:05:1}), provide
excellent bounds on the performance of iterative decoding techniques, to the
best of our knowledge, the best analytic performance bound in the case of MSA
decoding is still the bound given by Wiberg in his thesis based on the weight
distribution of a tree-like neighborhood of a vertex in a graph. A similar
bound was also derived by Lentmaier et
al.~\cite{Lentmaier:Truhachev:Costello:Zigangirov:04:1}. We derive the
equivalent bound for LP decoding of regular LDPC codes.

\section{Notation and Basics}

In this paper we are interested in binary LDPC codes where a binary LDPC code
$\code{C}$ of length $n$ is defined as the null-space of a sparse binary
parity-check matrix $\matr{H}$, i.e.~$\code{C} \defeq \{ \vx \in \GF{2}^n \ |
\ \matr{H} \vx^\tr = \vzero^\tr \}$. In particular, we focus on the case of
regular codes: an LDPC code $\code{C}$ is called $(J,K)$-regular if each
column of $\matr{H}$ has Hamming weight $J$ and each row has Hamming weight
$K$. The rate of a $(J,K)$-regular code is lower bounded by $1 - J/K$. To an
$M \times N$ parity-check matrix $\matr{H}$ we can naturally associate a
bipartite graph, the so-called Tanner graph $\graph{T}(\matr{H})$. This graph
contains two classes of nodes: variable nodes $\Vv$ and check nodes $\Vc$.
Both variable nodes and check nodes are identified with subsets of the
integers.  Variable nodes are denoted as $\Vv \defeq \{ 0, 1, \ldots, N-1 \}$
and check nodes are denoted as $\Vc \defeq \{0, 1, \ldots, M-1\}$. Whenever we
want to express that an integer belongs to the set of variable nodes we write
$i\in \Vv$; similarly, when an integer belongs to the set of check nodes we
write $j\in \Vc$. The Tanner graph $\graph{T}(\matr{H})$ contains an edge
$(i,j)$ between node $i\in \Vv$ and $j \in \Vc$ if and only if the entry
$h_{i,j}$ is non-zero. The set of neighbors of a node $i \in \Vv$ is denoted
as $\neighborhood{i}$; similarly, the set of neighbors of a node $j \in \Vc$
is denoted as $\neighborhood{j}$. In the following, $\set{E} \defeq \{ (i,j)
\in \Vv \times \Vc \ | \ i \in \Vv, \ j \in \neighborhood{i} \} = \{ (i,j) \in
\Vv \times \Vc \ | \ j \in \Vc, \ i \in \neighborhood{j} \}$ will be the set
of edges in the Tanner graph $\graph{T}(\matr{H})$. The convex hull of a set
$\set{A} \subseteq \R^n$ is denoted by $\convhull(\set{A})$. If $\set{A}$ is a
subset of $\GF{2}^n$ then $\convhull(\set{A})$ denotes the convex hull of the
set $\set{A}$ after $\set{A}$ has been canonically embedded in $\R^n$. The
inner product between vectors $\vect{x}$ and $\vect{y}$ is denoted as $\langle
\vect{x}, \vect{y} \rangle = \sum_l x_l y_l$. Finally, we define the set of
all binary vectors of length $K$ and even weight as $\code{B}^{(K)}$.

In the rest of this paper we assume that the all-zeros word was transmitted
--- an assumption without any essential loss of generality because we only
consider binary linear codes that are used for data transmission over a
binary-input output-symmetric channel. Given a received vector $\vy$ we define
the vector $\vlambda \defeq (\lambda_0, \lambda_1, \ldots, \lambda_{N-1})$ of
log-likelihood ratios by
\begin{align*}
  \lambda_i
    &= \log
         \left(
           \frac{P_{Y|X}(y_i|0)}
                {P_{Y|X}(y_i|1)}
         \right).
\end{align*}

\section{LP Decoding}

Maximum likelihood (ML) decoding may be cast as a linear program once we have
translated the problem into $\R^N$. To this end we embed the code into $\R^N$
by straightforward identification of $\F_2 = \{ 0, 1 \}$ with $\{0,1\} \subset
\R$. In other words, a code $\code{C}$ is identified with a subset of
$\{0,1\}^N \subset \R^N$.

\progbox{Maximum Likelihood Decoding}
{
\noindent { Minimize:} $\langle \vlambda,\vx\rangle$
 
\noindent { Subject to:} $\vx\in \convhull(\code{C})$
}

\noindent This description is usually not practical since the polytope
$\convhull(\code{C})$ is typically very hard to describe by hyperplanes (or as
a convex combination of points). Given a parity-check matrix $\matr{H}$, the
linear program is relaxed to~\cite{Feldman:03:1,
Feldman:Wainwright:Karger:05:1}

\progbox{LP Decoding}
{
\noindent { Minimize:} $\langle \vlambda,\vx\rangle$
 
\noindent { Subject to:} $\vx\in \fph{P}{H}$
}

\noindent Here, $\fph{P}{H}$ is the so-called fundamental
polytope~\cite{Feldman:03:1, Feldman:Wainwright:Karger:05:1,
Koetter:Vontobel:03:1, Vontobel:Koetter:05:1:subm} which is defined as
\begin{align*}
  \fph{P}{H}
    &\defeq
       \bigcap_{j = 0}^{M-1} 
         \convhull(\code{C}_j),
\end{align*}
where
\begin{align*}
  \code{C}_j
    &\defeq
       \code{C}_j(\matr{H})
     \defeq
       \left\{
         \vx \in \GF{2}^n 
           \ | \ \vect{h}_j \vx^\tr = 0
                 \text{ (mod $2$)} 
       \right\},
\end{align*}
where $\vect{h}_j$ is the $j$-th row of $\matr{H}$.

Since $\zero$ is always a feasible point, i.e. $\zero \in \fph{P}{H}$ holds,
zero is an upper bound on the value of the LP in LP decoding.  In fact, we can
turn this statement around by saying that whenever the value of the linear
program equals zero then the all-zeros codeword will be among the solutions to
the LP. Thus, motivated by the assumption that the all-zeros codeword was
transmitted, we focus our attention on showing that, under suitable
conditions, the value of the LP is zero which implies that the all-zeros
codeword will be found as a solution. For simplicity we only consider channels
where the channel output is a continuous random variable.  In this case a zero
value of the LP implies that the zero word is the unique solution with
probability one.  The main idea now is to show that the value of the dual
linear program is zero.  This technique, dubbed ``dual witness'' by Feldman et
al. in \cite{Feldman:Malkin:Stein:Servedio:Wainwright:04:1} will then imply
the correct decoding.

First, however, we need to establish the dual linear program. To this
end, for each $(i,j) \in \set{E}$, we associate the variable
$\tau_{i,j}$ with the edge between variable node $i$ and check node
$j$ in the Tanner graph $\graph{T}(\matr{H})$. In other words, we have
a variable $\tau_{i,j}$ if and only if the entry $h_{i,j}$ is
non-zero. For each $j \in \Vc$ we define the vector $\vtau_j$ that
collects all the variables $\{ \tau_{i,j} \}_{i \in
  \neighborhood{j}}$. Also, for each $j \in \Vc$, we associate the
variable $\theta_j$ with the check node $j$. We have\footnote{In the
  formal dual program the equality constraint
  $\sum_{j\in\neighborhood{i}} \tau_{i,j}=\lambda_i$ is an inequality
  ($\leq$).  However, there always exists a maximizing assignment of
  dual variables that satisfies this conditions with equality.}

\progbox{Dual LP}
{
\noindent Maximize:  $\sum_{j=0}^{M-1} \theta_j$
 
\noindent Subject to: $\theta_j \leq \langle \vx, \vtau_j \rangle$
          \quad\quad\quad\quad
          $\forall \ 
            j \in \Vc$, 
          \ \ 
          $\forall
            \vx \in \code{B}^{(K)}$

\noindent \hspace{1.5cm} 
          $\sum_{j\in\neighborhood{i}} \tau_{i,j}=\lambda_i$
          \quad\quad
          $\forall\ i\in\ \Vv$
}

The dual program has a number of nice interpretations. Any $\theta_j$
is bounded from above by zero and can only equal zero if the vector
$\vtau_j$ has minimal correlation with the all-zeros
codeword.\footnote{In a generalized LDPC code setting, the local code
  $\code{B}^{(K)}$ would have to be replaced by the corresponding
  code.} Thus the dual program will only get a zero value if we find
an assignment to $\tau_{i,j}$ such that the local all-zeros words are
among the ``best'' words for all $j$. We are constraint in setting the
$\tau_{i,j}$-values by the second equality constraint.

\section{MSA Decoding}

While MSA decoding is not the focus of interest in this paper, it turns out
that the MSA lies at the core of the proof technique that we will use. The MSA
is an algorithm that is being run until a predetermined criterion is
reached. With each edge in the graph we associate two messages: one message is
going towards the check-node and one is directed towards the variable
node. Let the two messages be denoted by $\mu_{i,j}$ and $\nu_{i,j}$,
respectively, where, as in the case of the single variable $\tau_{i,j}$ in the
section above, variables are only defined if the entry $h_{i,j}$ is
non-zero. The update rules of MSA are then

\progbox{Min-Sum Algorithm (MSA)}
{
  \noindent Initialize all variables $\nu_{i,j}$ to zero.
   
  \noindent 1) For all $(i,j)\in \set{E}$, let
    \begin{align*}
      \mu_{i,j}
        &:=
           \lambda_i
           +
           \sum_{j' \in \neighborhood{i} \setminus \{ j \}}
             \nu_{j',i}.
    \end{align*}
  
  \noindent 2) For all $(i,j) \in \set{E}$, let
    \begin{align*}
      \nu_{i,j}
        &:=
           \left(
             \prod_{i'\in \neighborhood{j} \setminus \{ i \}}
               \operatorname{sign}(\mu_{j,i'})
           \right) \\
        &\quad\quad\quad\quad
           \cdot
           \min
             \left\{
               \left|
                 \mu_{j,i'}
               \right|
               \, : \, 
               i'\in \neighborhood{j} \setminus \{ i \}
             \right\}.
    \end{align*}
}

\noindent Rather than the quantity $\nu_{i,j}$ we will consider its negative
value. Moreover, we keep track of the messages that were sent by message
numbers in the superscript. Thus we modify the MSA update equations as

\progbox{Modified Min-Sum Algorithm (modified MSA)}
{
  \noindent Initialize all variables $\nut{1}_{i,j}$ to zero.
  
  \noindent 
   
  \noindent 1) For all $(i,j) \in \set{E}$, let
  \begin{align*}
    \mut{s}_{i,j}
      &:=
         \lambda_i
         -
         \sum_{j' \in \neighborhood{i} \setminus \{ j \}}
           \nut{s}_{j',i}.
  \end{align*}

  \noindent 2) For all $(i,j) \in \set{E}$, let
    \begin{align*}
      \nut{s+1}_{i,j}
        &:= 
           -
           \left(
             \prod_{i' \in \neighborhood{j} \setminus \{ i \}}
               \operatorname{sign}(\mut{s}_{j,i'})
           \right) \\
        &\quad\quad\quad\quad
         \cdot
         \min
           \left\{
             \left|
               \mut{s}_{j,i'}
             \right| \, : \, i'\in \neighborhood{j} \setminus \{ i \}
           \right\}.
    \end{align*}
}

Clearly, the sign change leaves the algorithmic update steps essentially
unchanged. (Note that e.g. when all $\{ \mut{s}_{i,j} \}_{i \in
\neighborhood{j}}$ are non-negative then all $\{ \nut{s}_{i,j} \}_{i \in
\neighborhood{i}}$ will be non-positive.) Still, we may e.g.~write
$\mut{s}_{i,j} + \sum_{j' \in \neighborhood{i} \setminus \{ j \}}
\nut{s}_{j',i}=\lambda_i$ which more closely reflects the structure of the
dual program above.

We will need the notion of a computation tree (CT)~\cite{Wiberg:96}. We can
distinguish two types of CTs, rooted either at a variable node or at a check
node. Our CTs will be rooted at check nodes which is more natural when dealing
with the dual program.  A CT of depth $L$ consists of all nodes in the
universal cover of the Tanner graph that are reachable in $2L-1$ steps. In
particular, we will most of the time assume that the leaves in the CT are
variable nodes.

Assume we have run the MSA for $L$ iterations, corresponding to a CT of depth
$L$. For the moment let us also assume that the underlying graph has girth
larger than $4L$. Based on the iterations of the MSA and fixed CT root node
$j_0 \in \Vc$ we can assign values to the dual variables in the following way.

Was assign values to $\tau_{i,j}$ according to the distance of the edge
$(i,j)$ to the root node of the CT. So, if $(i,j)$ is at distance $2\ell+1$
from the root node $j_0$ then $\tau_{i,j}$ is assigned the value
$\mut{L-\ell}_{i,j}$ and if $(i,j)$ is at distance $2\ell+2$ from the root
node $j_0$ then $\tau_{i,j}$ is assigned the value
$\nut{L-\ell}_{i,j}$.\footnote{Edges incident to the root are said to be at
distance one. If the distance of the edge $(i,j)$ to the root $j_0$ is larger
than $2L$ then $\tau_{i,j} \defeq 0$.} Let us denote this assignment to
variables $\tau_{i,j}$ as $\vtau(j_0,L)$\footnote{The $j_0$ indicates that the
assignment is based on the CT rooted at node $j_0$.}. Note that the assignment
$\vtau(j_0, L)$ does not satisfy the constraints of the dual linear program,
i.e. itself it is not dual feasible. In particular, any edge of distance more
than $2L$ from the root is assigned the value $0$ and hence at any variable
node $i$ at distance more than $2L$ from the root we do not satisfy the
constraint
\begin{align*}
  \sum_{j\in\neighborhood{i}}
    \tau_{i,j} = \lambda_i,
\end{align*}
unless $\lambda_i$ happens to be $0$. However, we have the following lemma.

\begin{Lemma}
  \label{lemma:assignment:1}

  For each $j_0 \in \Vc$ let an assignment $\vtau(j_0, L)$ be given based on
  $L$ iterations of the MSA. The sum
  \begin{align*}
    \vtau(L)
      &\defeq \sum_{j_0 \in \Vc}
                \vtau(j_0, L)
  \end{align*}
  is a multiple of a dual feasible point. More precisely, for the number $T(L)
  \defeq \sum_{\ell=1}^L J \bigl[ (K-1)(J-1) \bigr]^{(\ell-1)}$ the vector
  \begin{align*}
    \frac{1}{T(L)} \vtau(L)
  \end{align*}
  is a dual feasible point.
\end{Lemma}

\begin{Proof}
  Each variable node $i \in \Vv$ is part of $\sum_{\ell=1}^LJ\bigl[ (K-1)(J-1)
  \bigr]^{(\ell-1)}$ CTs for different root nodes $j_0$ and so one can verify
  that we must have $\tau_{i,j}(L) = \sum_{j_0 \in \Vc} \tau_{i,j}(j_0, L) =
  \sum_{\ell=1}^{L} J \bigl[ (K-1)(J-1) \bigr]^{(\ell-1)}\lambda_i$. Using the
  abbreviation $T(L) \defeq \sum_{\ell=1}^L J \bigl[ (K-1)(J-1)
  \bigr]^{\ell-1}$ we see that
  \begin{align*}
    \frac{1}{T(L)} \vtau(L)
  \end{align*}
  is a dual feasible point.
\end{Proof}

The above lemma gives a structured way to derive dual feasible points for LP
decoding from the messages passed during the operation of the MSA. However,
these points are not very good since the overall assignment $\vtau(L)$ is
again dominated by the leaves of the CT with all the pertaining problems. The
problem becomes obvious when we write out the assignment $\vtau(L)$ as a
function of the MSA messages directly. If we perform $L$ steps of iterative
decoding, for any edge $(i,j) \in \set{E}$ we can write
\begin{align*}
  \tau_{i,j}(L)
    &= \mut{L}_{i,j}
       +(J-1) \left(
                \nut{L}_{i,j}
                +
                (K-1)\mut{L-1}_{i,j}
              \right) \\
    &\quad\ 
       +
       (J-1)^2(K-1)\left(
                     \nut{L-1}_{i,j}
                     + 
                     (K-1)\mut{L-2}_{i,j}
                   \right) \\
    &\quad\
       + \cdots \ .
\end{align*}
Written in form of a telescoping sum we get
\begin{align*}
  \tau_{i,j}(L)
    &= \mut{L}_{i,j}
       +
       (J-1)\Bigg(
              \nut{L}_{i,j}
              +
              (K-1)\bigg(\mut{L-1}_{i,j} \, + \\
    &\quad\ \       (J-1)
                    \Big(
                      \nut{L-1}_{i,j}
                      +
                      (K-1)
                      \big(
                        \mut{L-2}_{i,j}
                        +
                        \cdots
                      \big)
                    \Big)
                  \bigg)
            \Bigg).
\end{align*}
While the above sums show that the dual feasible point can be easily computed
alongside the MSA recursions it also shows the problem that messages
$\mut{\ell}_{i,j}$ and $\nut{\ell}_{i,j}$ are weighted exponentially more for small values of $\ell$.

We will have to attenuate the influence of the leaves in the CTs in order to
make interesting statements. To this end, let $\valpha$ be a vector with positive entries of
length $L$ and let a generalized assignment $\vtau(j_0, L, \valpha)$ to
dual variables be derived from $\vtau(j_0, L)$ by multiplying the message on
each edge at distance $2\ell+1$ or $2\ell+2$ by $\alpha_\ell$.\footnote{An
edge that is incident to a node $j$ is said to be at distance one from $j$;
$\alpha_0$ is set to one.} In other words, values assigned to edges at
distance three or four from the root node are multiplied with $\alpha_1$,
values at distance five and six are multiplied with $\alpha_2$ etc. Again we
can form the multiple of a dual feasible point as is shown in the next lemma.

\begin{Lemma}
  \label{lemma:assignment:2}

  For each $j_0 \in \Vc$ let an assignment $\vtau(j_0, L)$ be given based on
  $L$ iterations of the MSA. The sum
  \begin{align*}
    \vtau(L,\valpha)
      &\defeq \sum_{j_0 \in \Vc} \vtau(j_0, L, \valpha)
  \end{align*}
  is a multiple of a dual feasible point.
\end{Lemma}

\begin{Proof}
  Each variable node $i \in \Vv$ is part of $\sum_{\ell=1}^LJ\bigl[ (K-1)(J-1)
  \bigr]^{(\ell-1)}$ CTs for different root nodes $j_0$. Because all edges
  incident to a variable node are attenuated in the same way, one can verify
  that we must have $\tau_{i,j}(L, \valpha) = \sum_{j_0 \in \Vc}
  \tau_{i,j}(j_0, L, \valpha) = \sum_{\ell=1}^{L} \alpha_{\ell-1} J \bigl[
  (K-1)(J-1) \bigr]^{(\ell-1)}\lambda_i$. Using the abbreviation $T(L) \defeq
  \sum_{\ell=1}^L \alpha_{\ell-1} J \bigl[ (K-1)(J-1) \bigr]^{\ell-1}$ we see
  that
  \begin{align*}
    \frac{1}{T(L)} \vtau(L)
  \end{align*}
  is a dual feasible point.
\end{Proof}

Optimizing the vector $\valpha$ gives us some freedom and we want to choose
the vector $\valpha$ appropriately. First we have to learn more about the dual
feasible point that we construct in this way. While we kept the feasibility of
an assignment $\vtau(L,\valpha)$ by identically scaling the values
$\tau_{i,j}$ that are adjacent to a variable node in a CT, we scale values
$\tau_{i,j}$ that are adjacent to check nodes differently. Given a vector
$\valpha$, the dual feasible point may be easily computed together with the
messages of the MSA. To this end define a vector $\vbeta$ with components
$\beta_{\ell} \defeq \frac{\alpha_{\ell}}{\alpha_{\ell-1}}$. Writing again the
dual variable $\tau_{i,j}(L, \valpha)$ as functions of $\mut{\ell}_{i,j}$ and
$\nut{\ell}_{i,j}$ we get
\begin{align*}
  &
  \tau_{i,j}(L, \valpha) \\
    &= \mut{L}_{i,j}
       +(J-1)\left(
               \nut{L}_{i,j}
               +
               (K-1)\alpha_1 \mut{L-1}_{i,j}
             \right) \\
    &\quad\ 
       +
       (J-1)^2 (K-1) \left(
                       \alpha_1 \nut{L-1}_{i,j}
                       +
                       (K-1) \alpha_2 \mut{L-2}_{i,j}
                     \right) \\
    &\quad\ 
       +
       \cdots \ .
\end{align*}
Written in form of a telescoping sum we obtain
\begin{align*}
  &
  \tau_{i,j}(L, \valpha) \\
    &= \mut{L}_{i,j}
       +
       (J-1)\Bigg(
              \nut{L}_{i,j}
              +
              \beta_1(K-1)
              \bigg(\mut{L-1}_{i,j}
                \, + \\
     &\quad\ 
                (J-1)
                \Big(
                  \nut{L-1}_{i,j}
                  +
                  \beta_2(K-1)
                  \big(\mut{L-2}_{i,j}
                    +
                    \cdots
                  \big)
                \Big)
              \bigg)
            \Bigg).
\end{align*}

A particularly interesting choice for $\beta_\ell$ is $\beta_\ell \defeq
\frac{1}{K-1}$. The main reason for this choice is given in the following
lemma.

\begin{Lemma}
  \label{lemma:inner:product:1}
 
  Let $K >2$ and fix some $j \in \Vc$. Assume the MSA yields messages where
  $\mut{\ell}_{i,j}$ is positive for all $i \in \neighborhood{j}$ for some
  $\ell$. The inner product
  \begin{align*}
    \sum_{i \in \neighborhood{j}}
      b_i \left( \mut{\ell}_{i,j} + \nut{\ell+1}_{i,j} \right)
  \end{align*}
  is non-negative for all $\vb \in \code{B}^{(K)}$, in particular it is
  positive for all $\vb \in \code{B}^{(K)} \setminus \{ \vect{0}
  \}$.\footnote{We assume that the indices of $\vb$ are given by
  $\neighborhood{j}$.}
\end{Lemma}

\begin{Proof}
  Recall that $\nut{\ell}_{i,j}$ is negative for all $(i,j) \in \set{E}$ (this
  is in line with the modification of the MSA). One can easily verify the
  following fact about the vector containing $\mut{\ell}_{i,j} +
  \nut{\ell+1}_{i,j}$ for all $i \in \neighborhood{j}$: there is only one
  negative entry and the absolute value of this entry matches the absolute
  value of the smallest positive entry. The statement follows.
\end{Proof}

With the choice of $\alpha_i \defeq (K-1)^{-i}$, which results in
$\beta_i=\frac{1}{K-1}$, we get the following expression for the dual feasible
point
\begin{align*}
  \tau_{i,j}(L, \valpha)
    &= \mut{L}_{i,j}
       +
       (J-1)\bigg(
              \nut{L}_{i,j}
              + 
              \mut{L-1}_{i,j} \\
    &\quad\ 
       +
              (J-1)\Big(
                     \nut{L-1}_{i,j}
                     +
                     \mut{L-2}_{i,j}
                     +
                     (J-1)(\cdots)
                   \Big)
            \bigg)
\end{align*}
or
\begin{align*}
  \tau_{i,j}(L, \valpha)
    &= \mut{L}_{i,j}
       +
       (J-1) \left(
               \nut{L}_{i,j}
               +
               \mut{L-1}_{i,j}
             \right) \\
     &\quad\ 
       +
       (J-1)^2 \left(
                 \nut{L-1}_{i,j}
                 +
                 \mut{L-2}_{i,j}
               \right) \\
     &\quad\ 
       +
       \cdots
       +
       (J-1)^{L-1}
               \left(
                 \nut{2}_{i,j}
                 +
                 \mut{1}_{i,j}
               \right).
\end{align*}
We are still in a situation where $\mut{1}_{i,j}$ is weighted by a factor that
grows exponentially fast in $L$. However, we note that, once the MSA has
converged, $\mut{\ell}_{i,j}$ also grows exponentially fast in $\ell$ and this
offsets, to some extend, the exponential weighing of $\mut{1}_{i,j}$. In order
to exploit this fact more systematically we initialize the MSA's check to
variable messages $\nut{1}_{i,j}$, $(i,j) \in \set{E}$, with $-U$, where $U$
is a large enough positive number. With this initialization we can guarantee
(for $K > 2$) for all $(i,j) \in \set{E}$ that the value of $\mut{\ell}_{i,j}$
is strictly positive.\footnote{We may choose as any number greater than
$|\min(\lambda_i)/(J-2)|$.}  Thus we can apply
Lemma~\ref{lemma:inner:product:1}. It remains to offset the choice
$\nut{1}_{i,j}$ with $\mut{L}_{i,j}$.

To this end we consider a CT of depth $L$ rooted at check node $j_0$.
Consider the event $A_{j_0}$ that the all-zeros word on
this CT is more likely than any word that corresponds to a local
nonzero word assigned to the root node. \footnote{Event $A_{j_0}$ is
  defined on the CT without the change in initialization}.

\begin{Lemma}
  \label{lemma:assignment:3}

  Let $K > 2$ and assume the event $A_{j_0}$ is true. Moreover, assume that we
  initialize the MSA with check to variable messages $\nut{1}_{i,j} = -U$,
  $(i,j) \in \set{E}$, for a large enough number $U$. The inner product
  \begin{align*}
    \sum_{i \in \neighborhood{j}}
      b_i \left( \mut{L}_{i,j} + (J-1)^L \nut{1}_{i,j} \right)
  \end{align*}
  is non-negative for all $\vb \in \code{B}^{(K)}$, in particular it is
  non-negative  for all $\vb \in \code{B}^{(K)} \setminus \{ \vect{0}
  \}$.
\end{Lemma}

\begin{Proof}
  We exploit the fact that summaries sent by the MSA can be identified with
  cost differences of log-likelihood ratios.  Consider a message
  $\mut{L}_{i,j_0}$ on edge $(i,j_0)$.  This message may be written as
  $\mut{L}_{i,j_0}=\rho_i-(J-1)^L \nut{1}_{i,j_0}$ for some $\rho_i$.  Since
  the MSA propagates cost summaries along edges, we can interpret $\rho_i$ as
  the summary of the cost due to the $\lambda_i$ inside the subtree that
  emerges along the edge $(i,j_0)$. Similarly, $(J-1)^L \nut{1}_{i,j_0} $ is
  the cost contributed by the leaf nodes of this sub-tree. Here we use the
  fact that the minimal codeword which accounts for a one-assignment in edge
  $(i,j_0)$ contains exactly $(J-1)^L$ leaf nodes with a one-assignment.  But
  then the vector $(\mut{L}_{1,j_0}, \mut{L}_{2,j_0}, \ldots,
  \mut{L}_{|\neighborhood{j_0}|, j_0}) + (J-1)^L(\nut{1}_{1,j_0},
  \nut{1}_{2,j_0}, \ldots, \nut{1}_{|\neighborhood{j_0}|,j_0})$ equals the
  vector $\vrho \defeq (\rho_1, \rho_2, \ldots,
  \rho_{|\neighborhood{j_0}|})$. The event $A_{j_0}$ is true only if the inner
  product $\langle \vrho,\vb \rangle$ is positive for all $\vb \in
  \code{B}^{(K)} \setminus \{ \vect{0}\}$. Hence event $A_{j_0}$ implies the
  claim of the lemma.
\end{Proof}

Let $\vtau^*$ be the averaged assignment to the dual variables obtained from
the MSA messages with $\nut{1}_{i,j}$ set to $-U$. 
Lemmas~\ref{lemma:inner:product:1} and~\ref{lemma:assignment:3} imply that the
sum, 
\begin{align*}
  \sum_{i \in \neighborhood{j}}
    b_i \tau^*_{i, j_0}
\end{align*}
has a non-negative value for any $\vb \in \code{B}^{(K)}$, and, in particular,
the value equals zero for $\vb=\zero$. It follows that $\theta_{j_0}$ in the
dual LP can be chosen as zero.

For each check node $j$ for which event $A_j$ is true we can be sure
that the correlation of any codeword in $\code{B}^{(K)}$ with
$\vtau_j^{*}$ is non-negative.
If we can be sure that the event $A_j$ is true for all check nodes we
would, thus, have exhibited a dual witness for the optimality of the
all-zeros codeword. We have to estimate the probability of the event
$A_j$ and set it in relation to the number of checks in the graph
$\graph{T}(\matr{H})$. In order to estimate the latter we employ a
result by Gallager \cite{Gallager:63} that guarantees the existence of
$(J,K)$-regular graphs in which we can conduct $L$ steps of MSA
decoding without closing any cycles provided that $L$ satisfies
\begin{align}
  L
    &\leq
       \frac{\log(N)}
            {2\log((J-1)(K-1))}
       - \kappa
              \label{eq:girth}
\end{align}
where the term $\kappa$ in this expression is independent of $N$.

Finally, we can estimate the probability of the event $A_j$ from the known
weight distribution of the code on the CT provided the
underlying graph has girth at least $4L$.  The minimal codewords have
weight  $2
(1 + (J-1) + (J-1)^2 + \cdots (J-1)^{L-1})$ and there are a total of
\begin{align*}
  &
  \frac{K(K-1)}{2}\cdot (K-1)^{2(J-1)}\cdot(K-1)^{2(J-1)^2} \\
  &\quad\quad\ \cdots (K-1)^{2(J-1)^{L-1}} \\
  &\quad
     = K/2 (K-1)^{2(1+(J-1)+(J-1)^2+\ldots (J-1)^{L-1})}
\end{align*}
minimal-tree codewords. Based on a union bound we thus get an
expression
\begin{align}
  P(A_j)
    &< \frac{K}{2} \big( (K-1) \gamma \big)^{2\frac{(J-1)^L-1}{J-2}}
  \label{eq:aj}
\end{align}
which means that $P(A_j)$ decreases doubly exponentially in $L$ if the
Bhattacharyya parameter $\gamma$ satisfies $\gamma < \frac{1}{K-1}$.

Thus we have proved the following theorem:

\begin{Theorem}
  Let a sequence of $(J,K)$-regular LDPC codes be given that satisfies
  equation \eqref{eq:girth}.  Under LP decoding this sequence achieves an
  arbitrarily small probability of error on any memoryless channel for which
  the Bhattacharyya parameter $\gamma$ satisfies $\gamma \leq \frac{1}{K-1}$.
  For such a channel the word error probability $P_W$ decreases as
  \begin{align*}
    P_W
      &< \eta_1 2^{-\eta_2 N^{\frac{\log(J-1)}{2\log((J-1)(K-1))}}}
  \end{align*}
  for some positive parameters $\eta_1$ and $\eta_2$.
\end{Theorem}

\begin{Proof}
  Most of the proof is contained in the arguments leading up to the
  theorem. In order to see the explicit form of the word error rate we employ
  a union bound for the $N\frac{J}{K}$ check nodes combining \eqref{eq:girth}
  and \eqref{eq:aj}. We find that the word error rate is bounded by
  \begin{align*}
    P_W
      &< \frac{N J}{2} 
         \big(
           (K-1) \gamma 
         \big)
         ^{2\frac{(J-1)^{\frac{\log(N)}{2\log((J-1)(K-1))} - \kappa}-1}{J-2}},
  \end{align*}
  where $\kappa$ does not depend on $N$. The statement of the theorem is
  obtained by simplifying this expression.
\end{Proof}

\mbox{}
\vfill  
\newpage

We conclude this paper with an intriguing observation concerning the AWGN
channel. In \cite{Vontobel:Koetter:05:1:subm} it is proved that no
$(J,K)$-regular LDPC code can achieve an error probability behavior better
than  $P_W\geq \eta_3 2^{-\eta_4
N^{\frac{2\log(J-1)}{\log((J-1)(K-1))}}}$ for constants $\eta_3$ and $\eta_4$
that are independent on $N$. The result of the theorem thus shows that there
exist sequences of LDPC codes whose error probability behavior under LP
decoding is boxed in as a function of $N$ between:
\begin{align*}
  \eta_3 2^{-\eta_4 N^{\frac{2\log(J-1)}{\log((J-1)(K-1))}}}
    &\leq P_w 
     \leq \eta_1 
          2^{-\eta_2 N^{\frac{\log(J-1)}{2\log((J-1)(K-1))}}}
\end{align*}


\begin{thebibliography}{10}

\bibitem{Wiberg:96}
N.~Wiberg, {\em Codes and Decoding on General Graphs}.
\newblock PhD thesis, Link\"oping University, Sweden, 1996.

\bibitem{Luby:Mitzenmacher:Shokrollahi:Spielman:Stemann:97:1}
M.~G. Luby, M.~Mitzenmacher, M.~A. Shokrollahi, D.~A. Spielman, and V.~Stemann,
  ``Practical loss-resilient codes,'' in {\em Proc.~29th Annual ACM Symp.~on
  Theory of Computing}, pp.~150--159, 1997.

\bibitem{Luby:Mitzenmacher:Shokrollahi:98:1}
M.~G. Luby, M.~Mitzenmacher, and M.~A. Shokrollahi, ``Analysis of random
  processes via and-or tree evaluation,'' in {\em Proc.~9th Annual ACM-SIAM
  Symp.~on Discrete Algorithms}, pp.~364--373, 1998.

\bibitem{Richardson:Urbanke:01:2}
T.~J. Richardson and R.~L. Urbanke, ``The capacity of low-density parity-check
  codes under message-passing decoding,'' {\em IEEE Trans.\ on Inform.\
  Theory}, vol.~IT--47, no.~2, pp.~599--618, 2001.

\bibitem{Lentmaier:Truhachev:Costello:Zigangirov:04:1}
M.~Lentmaier, D.~V. Truhachev, D.~J. {Costello, Jr.}, and K.~Zigangirov, ``On
  the block error probability of iteratively decoded {LDPC} codes,'' in {\em
  5th ITG Conference on Source and Channel Coding}, (Erlangen, Germany),
  Jan.~14-16 2004.

\bibitem{Jin:Richardson:05:1}
H.~Jin and T.~Richardson, ``Block error iterative decoding capacity for {LDPC}
  codes,'' in {\em Proc.\ IEEE Intern.\ Symp.\ on Inform.\ Theory}, (Adelaide,
  Australia), pp.~52 -- 56, Sep.~4--9 2005.

\bibitem{Feldman:03:1}
J.~Feldman, {\em Decoding Error-Correcting Codes via Linear Programming}.
\newblock PhD thesis, Massachusetts Institute of Technology, Cambridge, MA,
  2003.
\newblock Available online under
  \verb+http://www.columbia.edu/+ \verb+~jf2189/pubs.html+.

\bibitem{Feldman:Wainwright:Karger:05:1}
J.~Feldman, M.~J. Wainwright, and D.~R. Karger, ``Using linear programming to
  decode binary linear codes,'' {\em IEEE Trans.\ on Inform.\ Theory},
  vol.~IT--51, no.~3, pp.~954--972, 2005.

\bibitem{Koetter:Vontobel:03:1}
R.~Koetter and P.~O. Vontobel, ``Graph covers and iterative decoding of
  {f}inite-length codes,'' in {\em Proc.\ 3rd Intern.~Symp.~on Turbo Codes and
  Related Topics}, (Brest, France), pp.~75--82, Sept.~1--5 2003.

\bibitem{Vontobel:Koetter:05:1:subm}
P.~O. Vontobel and R.~Koetter, ``Graph-cover decoding and {f}inite-length
  analysis of message-passing iterative decoding of {LDPC} codes,'' {\em
  submitted to IEEE Trans.\ Inform.\ Theory, available online under
  \verb+http://+ \verb+www.arxiv.org/abs/cs.IT/0512078+}, Dec. 2005.

\bibitem{Feldman:Malkin:Stein:Servedio:Wainwright:04:1}
J.~Feldman, T.~Malkin, C.~Stein, R.~A. Servedio, and M.~J. Wainwright, ``{LP}
  decoding corrects a constant fraction of errors,'' in {\em Proc.\ IEEE
  Intern.\ Symp.\ on Inform.\ Theory}, (Chicago, IL, USA), p.~68, June 27--July
  2 2004.

\bibitem{Gallager:63}
R.~G. Gallager, {\em Low-Density Parity-Check Codes}.
\newblock M.I.T. Press, Cambridge, MA, 1963.
\newblock Available online under
  \verb+http://web.mit.edu/+ \verb+gallager/www/pages/ldpc.pdf+.

\end{thebibliography}
\end{document}